\documentclass[12pt,cls,onecolumn]{IEEEtran}
\usepackage{graphicx,amsmath,amssymb,epsfig, amsfonts, cite, latexsym, cuted, multicol, multirow, subfigure, stfloats, array, tabularx}
\usepackage{subeqnarray,tabularx}
\usepackage{color}
\usepackage{setspace}
\usepackage{anysize}

\begin{document}

\title{Clarifying the Role of Distance in Friendships on Twitter: Discovery of a Double Power-Law Relationship}
\author{\large Won-Yong Shin, Jaehee Cho, and
Andr{\'e} M. Everett
\\
\thanks{This research was supported by the Basic Science Research
Program through the National Research Foundation of Korea (NRF)
funded by the Ministry of Education (2014R1A1A2054577).}
\thanks{W.-Y. Shin is with the Department of Computer Science and
Engineering, Dankook University, Yongin 448-701, Republic of Korea
(E-mail: wyshin@dankook.ac.kr).}
\thanks{J. Cho is with the Department of Business Administration, Kwangwoon University, Seoul 139-701, Republic of Korea (E-mail: mis1@kw.ac.kr)}
\thanks{A. M. Everett is with the Department of Management, University of Otago,
Dunedin 9054, New Zealand (E-mail: andre.everett@otago.ac.nz).}
} \maketitle


\markboth{ACM SIGSPATIAL'15} {Shin {\em et al.}: Clarifying the
Role of Distance in Friendships on Twitter: Discovery of a Double
Power-Law Relationship}


\newtheorem{definition}{Definition}
\newtheorem{theorem}{Theorem}
\newtheorem{lemma}{Lemma}
\newtheorem{example}{Example}
\newtheorem{corollary}{Corollary}
\newtheorem{proposition}{Proposition}
\newtheorem{conjecture}{Conjecture}
\newtheorem{remark}{Remark}

\def \diag{\operatornamewithlimits{diag}}
\def \min{\operatornamewithlimits{min}}
\def \max{\operatornamewithlimits{max}}
\def \log{\operatorname{log}}
\def \max{\operatorname{max}}
\def \rank{\operatorname{rank}}
\def \out{\operatorname{out}}
\def \exp{\operatorname{exp}}
\def \arg{\operatorname{arg}}
\def \E{\operatorname{E}}
\def \tr{\operatorname{tr}}
\def \SNR{\operatorname{SNR}}
\def \dB{\operatorname{dB}}
\def \ln{\operatorname{ln}}

\def \bmat{ \begin{bmatrix} }
\def \emat{ \end{bmatrix} }

\def \be {\begin{eqnarray}}
\def \ee {\end{eqnarray}}
\def \ben {\begin{eqnarray*}}
\def \een {\end{eqnarray*}}

\begin{abstract}
This study analyzes friendships in online social networks
involving geographic distance with a geo-referenced Twitter
dataset, which provides the exact distance between corresponding
users. We start by introducing a strong definition of ``{\em
friend}" on Twitter, requiring {\em bidirectional communication}.
Next, by utilizing {\em geo-tagged mentions} delivered by users to
determine their locations, we introduce a two-stage distance
estimation algorithm. As our main contribution, our study provides
the following newly-discovered friendship degree related to the
issue of space: The number of friends according to distance
follows a {\em double power-law} (i.e., a {\em double Pareto law})
distribution, indicating that the probability of befriending a
particular Twitter user is significantly reduced beyond a certain
geographic distance between users, termed the {\em separation
point}. Our analysis provides much more fine-grained social ties
in space, compared to the conventional results showing a
homogeneous power-law with distance.
\end{abstract}


\newpage

\section{Introduction}


To understand the nature of friendships online with respect to
geographic distance, some efforts have originally focused on
users' online profiles that include their city of residence.
In~\cite{Liben-Nowell}, experimental results based on the
LiveJournal social network demonstrated a close relationship
between geographic distance and probability distribution of
friendship, where the probability of befriending a particular user
on LiveJournal is inversely proportional to the positive power of
the number of closer users.
However, the geographic location points only to the location of
users at a {\em city scale}. For this reason, the friendship
degree distribution contains a background probability that is
independent of geography due to the city-scale resolution. As
follow-up studies, using the data collected from
Facebook~\cite{Backstrom} and three popular online location-based
social networks (LBSNs)~\cite{Scellato}, it was found that the
probability distribution of friendship as a function of distance
also closely follows a single power-law but represents some
heterogeneous features. More precisely, it is observed
in~\cite{Backstrom} that the corresponding curve has two regions
according to the population density, indicating that it is flatter
at shorter distances---a small fraction of Facebook users who
entered their home addresses were used. In~\cite{Scellato}, the
probability of friendship with distance was shown to present noisy
patterns such as an almost flatness in a certain range---the home
location of each user was defined as the place with the largest
number of check-ins. Contrary
to~\cite{Liben-Nowell,Backstrom,Scellato}, based on the data
collected from the Tuenti social network, it was found
in~\cite{Kaltenbrunner} that social interactions online are only
weakly affected by spatial proximity, with other factors
dominating.

Alternatively, there is extensive and growing interest among
researchers to understand a variety of social behaviors through
geo-tagged
tweets~\cite{Takhteyev,Kulshrestha,Alowibdi,Lee,Hawelka,Jurdak}.
The volume of geo-located Twitter has grown constantly and now
forms an invaluable register for understanding human behavior and
modelling the way people interact in space. In~\cite{Takhteyev},
along with geo-locations for collected tweets, analysis included
how geo-related factors such as physical distance, frequency of
air travel, national boundaries, and language differences affect
formation of social ties on Twitter. In~\cite{Kulshrestha}, it was
found that the geo-locations of Twitter users across different
countries considerably impact their participation in Twitter and
their connectivity with other users.
New approaches based on geo-tagged tweets were also proposed to
find top vacation spots for a particular holiday by applying
indexing, spatio-temporal querying, and machine learning
techniques~\cite{Alowibdi} and to detect unusual geo-social events
by measuring geographical regularities of crowd
behaviors~\cite{Lee}. Additionally, owing to the location
information from geo-tagged tweets, there has been a steady push
to understand individual human mobility~\cite{Hawelka,Jurdak},
which is of fundamental importance for many applications. Recent
effort has focused on the studies of human mobility using tracking
technologies such as mobile phones, GPS receivers, WiFi logging,
Bluetooth, and RFID devices as well as LBSN check-in
data~\cite{Cho}, but these technologies involve privacy concerns
or data access restrictions. On the other hand, geo-tagged tweets
can capture much richer features of human
mobility~\cite{Hawelka,Jurdak}.


In our work, we utilize {\em geo-tagged mentions} on Twitter, sent
by users, to identify their exact location information. A
`mention' in Twitter consists of inclusion of ``@username"
anywhere in the body of tweets. From the fact that we tend to
interact offline with people living very near to us, we derive as
a natural extension the question whether geography and social
relationships are inextricably intertwined on Twitter. Our
research is interested in how a pair of users interacts through
geo-tagged mentions.

As people normally spend a substantial amount of time online, data
regarding these two dimensions (i.e., geography and online social
relationships) are becoming increasingly precise, thus motivating
us to build more reliable models to describe social
interactions~\cite{Liben-Nowell,Backstrom,Scellato}. This paper
goes beyond past research to determine how friendship patterns are
geographically represented by Twitter, analyzing a single-source
dataset that contains a huge number of geo-tagged mentions from
users in i) the state of California in the United States (US) and
Los Angeles (the most populous city in the state) and ii) the
United Kingdom (UK) and London (the most populous city in the UK).
These two location sets were selected as demographically
comparable, yet distinct and geographically separated, leading
adopters of Twitter with sufficient data to enable meaningful
comparative analysis for our intentionally exploratory study. We
propose and apply the following framework, which establishes a
much more accurate friendship degree on Twitter, and a method to
enable analysis based on geographic distance:

\begin{itemize}
\item To fully take into account the intensity of communication
between users, we start our analysis by introducing a rather
strong definition of ``{\em friend}" on Twitter, i.e., a
definition of {\em bidirectional friendship}, instead of
na\"{\i}vely considering the set of followers and followees
(unidirectional terms). This definition requires bidirectional
communication within a designated time frame or creating a
friendship.

\item By showing that almost all Twitter users are likely to post
consecutive tweets in the static mode (i.e., no movement mode), we
propose a two-stage distance estimation method, where the
geographic distance between two befriended users based on our
definition of bidirectional friendship is estimated by
sequentially measuring the two senders' locations.
\end{itemize}

We would like to synthetically analyze how the geographic distance
between Twitter users affects their interaction, based on our new
framework. Our main results are summarized as follows:

\begin{itemize}
\item We characterize a newly-discovered probability distribution
of the number of friends according to {\em geographic distance},
which does not follow a homogenous power-law but, instead, a {\em
double power-law} (i.e., a {\em double Pareto law}).

\item From this new finding, we identify not only two
fundamentally {\em separate regimes}, which are characterized by
two different power-laws in the distribution, but also the {\em
separation point} between these regimes.
\end{itemize}

We refer to our full paper~\cite{Shin} for more detailed
description and all the rigorous steps.

\section{Dataset}

We use a dataset collected via Twitter Streaming API. The dataset
consists of a huge amount of geo-tagged mentions recorded from
Twitter users from September 22, 2014 to October 23, 2014 (about
one month) in the following four regions: California, Los Angeles,
UK, and London.
Note that this short-term (one month) dataset is sufficient to
examine how closely one user has recently interacted with another
online. In this dataset, each mention record has a geo-tag and a
timestamp indicating from where, when, and by whom the mention was
sent. Based on this information, we are able to construct a user's
location history denoted by a sequence $L=(x_{ki},y_{ki},t_i)$,
where $x_{ki}$ and $y_{ki}$ are the $x-$ and $y-$ coordinates of
User $k$ at time $t_i$, respectively. The location information
provided by the geo-tag is denoted by latitude and longitude,
which are measured in degrees, minutes, and seconds. Each mention
on Twitter contains a number of entities that are distinguished by
their attributed fields. For data analysis, we adopted the
following five essential fields from the metadata of mentions:

\begin{itemize}
\item {\em user\textunderscore id\textunderscore str}: string
representation of the sender ID \item {\em in\textunderscore
reply\textunderscore to\textunderscore user\textunderscore
id\textunderscore str}: string representation of the receiver ID
\item {\em lat}: latitude of the sender \item {\em lon}: longitude
of the sender \item {\em created\textunderscore at}: UTC/GMT time
when the mention is delivered, i.e., the timestamp
\end{itemize}

\section{Research Methodology} \label{SEC:method}

We start by introducing the following definition of
``bidirectional friendship" on Twitter.
\begin{definition}
If two users send/receive mentions to/from each other (i.e.,
bidirectional personal communication occurs) within a designated
amount of time, then they form a bidirectional friendship with
each other.
\end{definition}

Note that our definition differs from the conventional definition
of ``friend" on Twitter, which is referred to as a followee and
thus represents a {\em unidirectional} relation. This strong
definition enables exclusion of {\em inactive friends} who have
been out of contact online for a long designated amount of time
(e.g., about one month in our work) and to count the number of
{\em active friends} who have recently communicated with each
other.

Now, let us characterize the friendship degree of individuals
regarding geography by analyzing their sequences
$L=(x_{ki},y_{ki},t_i)$ of geo-tagged mentions, where only the
senders' location information is recorded. We propose a two-stage
distance estimation method, where the geographic distance between
two befriended users is estimated by sequentially measuring the
two senders' locations. We first focus on the time interval
between the following two events for a befriended pair: a mention
and its {\em replied} mention at the next closest time. We count
only the events with a time duration between a mention and its
replied mention, or inter-mention interval, of {\em less than one
hour} to exclude certain inaccurate location information that may
occur due to users' movements. We next consider the instance for
which User $u$, originally placed at $(x_{u0},y_{u0},t_0)$, sent a
mention to User $v$ at $(x_{v0},y_{v0},t_0)$, and then received a
replied mention at the location $(x_{u1},y_{u1},t_1)$ from User
$v$ placed at $(x_{v1},y_{v1},t_1)$. From these two consecutive
mention events, it is possible to estimate the geographic distance
based on the two sequences $(x_{u0},y_{u0},t_0)$ and
$(x_{v1},y_{v1},t_1)$. In our framework, by assuming that the
Earth is spherical, we deal with the shortest path between two
users' locations measured along the surface of the Earth. Then,
the distance between two locations on the Earth's surface can be
computed according to the spherical law of cosines, which gives a
well-conditioned result of the estimated distance down to
distances as small as around 1 meter. The estimated distance for
one pair is finally obtained by taking the average of all distance
values computed over the available inter-mention intervals, each
of which is less than one hour. While the estimated distance may
differ from the actual distance between Users $u$ and $v$ at time
$t_1$, it is worth noting that people tend to send/receive
multiple consecutive tweets from the same location to convey a
series of ideas~\cite{Jurdak}. Our supplementary experiments also
demonstrate that most of the Twitter users (approximately 90\%) in
the four regions under consideration are likely to post
consecutive tweets in the {\em static} mode whose average velocity
ranges from 0 to 2 km/h. Although the inter-tweet interval may
show a different pattern from that of the inter-mention interval,
we believe that our demonstration is sufficient to support our
analysis methodology.
%


\section{Analysis Results}

\begin{figure}[t!]
\centering{%
\subfigure[California]{%
\epsfxsize=0.49\textwidth \leavevmode \label{FIG:Distance_CA}
\epsffile{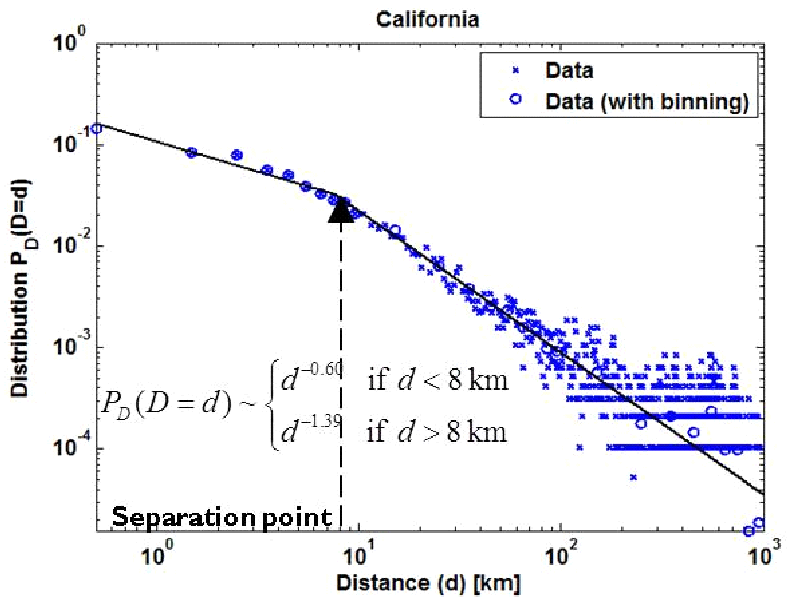}}\vspace{.0cm}
\subfigure[Los Angeles]{%
\epsfxsize=0.49\textwidth \leavevmode \label{FIG:Distance_LA}
\epsffile{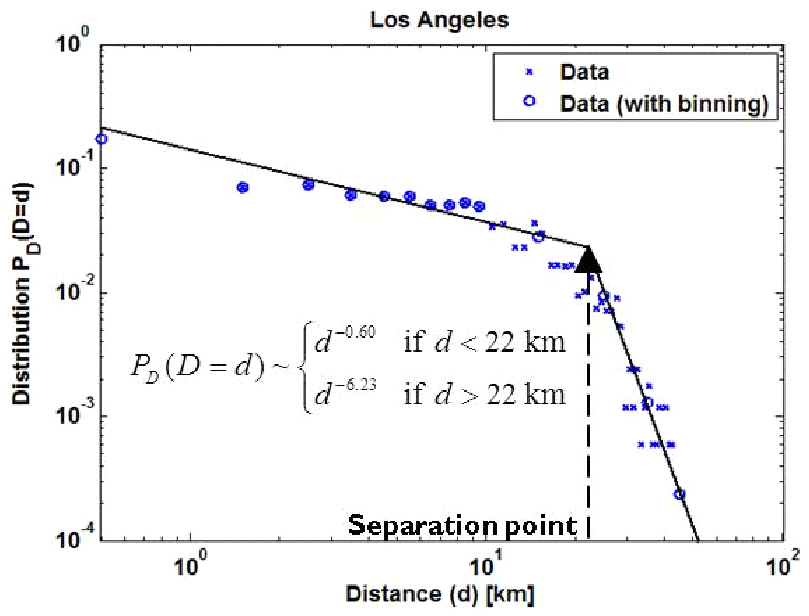}}\vspace{.0cm}
\subfigure[UK]{%
\epsfxsize=0.49\textwidth \leavevmode \label{FIG:Distance_UK}
\epsffile{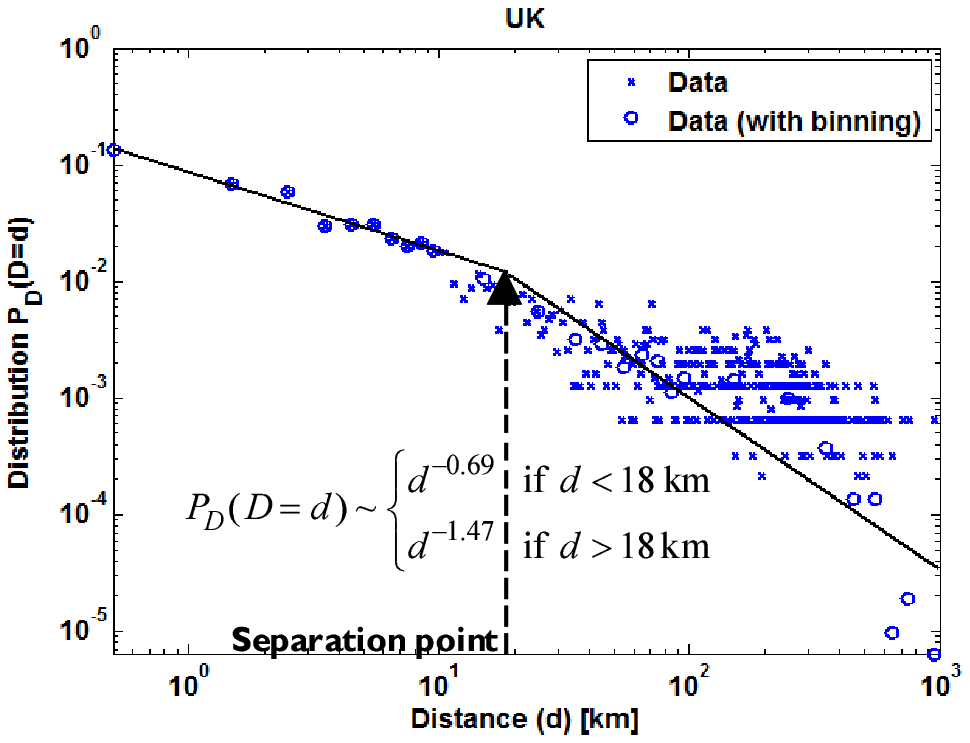}}\vspace{.0cm}
\subfigure[London]{%
\epsfxsize=0.49\textwidth \leavevmode \label{FIG:Distance_London}
\epsffile{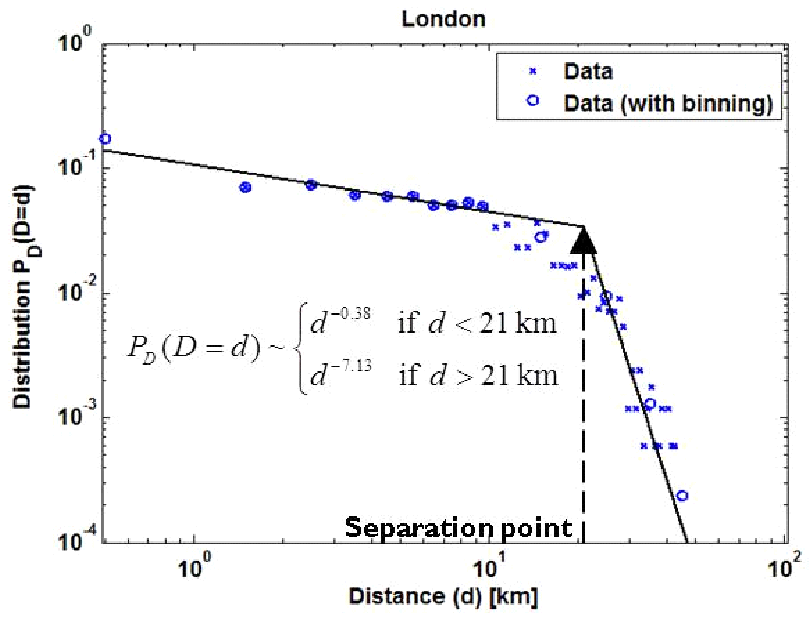}} } \caption{Probability distribution
$P_D(D=d)$ of the number of friends with respect to distance
(log-log plot).} \label{fig:distance}
\end{figure}

Using bidirectional mentions in Section~\ref{SEC:method}, we
characterize the probability distribution $P_D(D=d)$ of the number
of friends according to the distance $d$, where $d$ [km] is the
geographic distance between a user and his/her friend. Unlike the
earlier work
in~\cite{Liben-Nowell,Backstrom,Scellato,Kaltenbrunner}, the
heterogeneous shape of $P_D(D=d)$ for the entire interval cannot
be captured by a single commonly-used statistical function such as
a homogeneous power-law using the approach of parametric fitting.
Interestingly, we observe that for the distance $d\in
[d_{\text{min}}, d_{\text{max}}]$, $P_D(D=d)$ can be described as
a {\em double power-law} distribution, which is given below:
\begin{eqnarray}
P_D(D=d)\sim \left\{\begin{array}{lll} d^{-\gamma_1} &\textrm{if
$d_{\text{min}}\le d<d_s$ (intra-city regime)}
\\ d^{-\gamma_2} &\textrm{if $d_s\le d\le d_{\text{max}}$ (inter-city regime),}
\end{array}\right. \nonumber
\end{eqnarray}
where $\gamma_1$ and $\gamma_2$ denote the exponents for each
individual power-law and $d_s$ is the {\em separation point}. This
finding indicates that the friendship degree can be composed of
two separate regimes characterized by two different power-laws,
termed the {\em intra-city} and {\em inter-city} regimes.
Figure~\ref{fig:distance} shows the log-log plot of the
distribution $P_D(D=d)$ from empirical data, logarithmically
binned data, and fitting function, where the fitting is applied to
the binned data. As depicted in the figure, statistical noise
exists in the tail for large $d$, which can be eliminated by
applying logarithmic binning.\footnote{It is verified that this
binning procedure does not fundamentally change the underlying
power-law exponent of $P_D(D=d)$.} We use the traditional least
squares estimation to obtain the fitting function.\footnote{Using
maximum likelihood estimation to fit a mixture function (e.g., a
double power-law function) is not easy to implement and the
performance of a mixture function has not been well understood.}

Unlike the earlier studies that do not capture the friendship
patterns in the intra-city regime, our analysis exhibits two
distinguishable features with respect to distance. More
specifically, in each regime, the following interesting
observations are made:

\begin{itemize}
\item In the intra-city regime, $P_D(D=d)$ decays slowly with
distance $d$, which means that geographic proximity weakly affects
the number of intra-city friends with which one user interacts.
That is, in this regime, the geographic distance is less relevant
for determining the number of friends. This finding reveals that
more active Twitter users tend to preferentially interact over
{\em short-distance} connections.

\item In the inter-city regime, $P_D(D=d)$ depends strongly on the
geographic distance, where there exists a sharp transition in the
distribution $P_D(D=d)$ beyond the separation point $d_s$. Thus,
{\em long-distance} communication is made occasionally.
\end{itemize}

The above argument stems from the fact that the separation point
$d_s$ is closely related to the length and width of the city in
which a user resides. From these observations, we may conclude
that, within a given period, the individual is much more likely to
contact online mostly friends who are in location-based
communities that range from the local neighborhood, suburb,
village, or town up to the city level. In addition, the following
interesting comparisons are performed according to types of
regions:

\begin{itemize}
\item {\bf Comparison between the city-scale and
state-scale/country-scale results}: We observe that $d_s$ in
populous metropolitan areas is greater than that in larger regions
that include local small towns (such as at the state or country
level). For example, from Figures~1(a) and 1(b), we see that $d_s$
is 8 km and 22 km in California and Los Angeles, respectively.
From Figures 1(c) and 1(d), the same trend is observed by
comparing the results for the UK and London (18 km and 21 km,
respectively). This finding reveals that Twitter users in populous
metropolitan areas (e.g., Los Angeles and London) have a stronger
tendency to contact friends on Twitter who are geographically away
from their location (i.e., interacting over long-distance
connections). This is because the average size (referred to as the
land area) of the considered metropolitan cities is relatively
bigger than that of larger regions including small towns. It is
also seen that the exponent in the inter-city regimes (i.e.,
$\gamma_2$) in metropolitan areas is significantly higher than
that in larger regions. Unlike the state-scale/country-scale
results, this finding implies that $P_D(D=d)$ sharply drops off
beyond $d_s$ in huge metropolitan areas.

\item {\bf Comparison between the results in the two cities}: From
Figures~1(b) and 1(d), one can see that $\gamma_1$ is 0.60 and
0.38 and $\gamma_2$ is 6.23 and 7.13 in Los Angeles and London,
respectively. Thus, in the intra-city regime, the geographic
distance is less relevant in London for determining the number of
friends. However, in the inter-city regime, $P_D(D=d)$ in London
shows a bit steeper decline.
\end{itemize}


\end{document}